\documentclass[showpacs,preprintnumbers,amsmath,amssymb,aps, prd]{revtex4}
\usepackage{graphicx}
\usepackage{grffile}
\usepackage{mathrsfs, mathtools}
\usepackage{caption}
\usepackage{float}
\usepackage{xcolor}
\usepackage{dcolumn}
\usepackage{bm}
\usepackage{graphicx,subfigure,epsfig}
\usepackage{multirow}
\newcommand{\lagr}{\mathscr{L}}

\def\MN{^{\mu\nu}}
\def\mN{_\mu^\nu}

\def\d{\partial}
\def\ep{\epsilon}
\begin{document}
\title{Photonsphere, shadow, quasinormal modes, and greybody bounds of non-rotating Simpson-Visser black hole}
\author{Sohan Kumar Jha}
\email{sohan00slg@gmail.com}
\affiliation{Chandernagore College, Chandernagore, Hooghly, West
Bengal, India}

\date{\today}
\begin{abstract}
\begin{center}
Abstract
\end{center}
In this manuscript, we study photonsphere, shadow, quasinormal modes, Hawking temperature, and greybody bounds of a non-rotating Simpson-Visser black hole which is a regular black hole. We observe that though the radius of the photonsphere does depend on the Simpson-Visser parameter $\alpha$, the shadow radius is independent of it. The shadow radius is found to be equal to that for Schwarzschild black hole. We, then, study quasinormal frequencies of the Simpson-Visser black hole for scalar and electromagnetic perturbations with the help of $6$th order WKB method. We tabulate values of quasinormal frequencies for various values of $\alpha$, angular momentum $\ell$, and overtone number $n$. We also graphically show the dependence of real and imaginary parts of quasinormal frequency on $\alpha$ and $\ell$. Additionally, We study the convergence of the WKB method for various values of pair $(n,\ell)$. Finally, we shed light on the dependence of the Hawking temperature on the parameter $\alpha$ and the dependence of greybody bounds on $\alpha$ and $\ell$.
\end{abstract}
\maketitle
\section{Introduction}
Black holes(BHs) are one of the most fascinating objects in the Universe. A great deal of research has gone into studying various aspects of BHs. It was the General theory of relativity(GTR) proposed by Einstein that gave rise to the very idea of BHs \cite{EINSTEIN}. BHs that are derived from GTR such as Schwarzschild black holes or Reissner-Nordstrom(RN) black holes have two singularities: one is coordinate singularity called event horizon and the other is essential singularity at $r=0$. It is because of the presence of essential singularity that curvature invariants diverge and geodesics become incomplete. We can avoid this singularity in GTR if, in the vicinity of BHs, the strong energy condition is broken. BHs having event horizon but no essential singularity are called regular black holes(RBHs). For RBHs, curvature invariants are finite everywhere. Two different approaches can be used to generate RBHs solutions. One approach is to consider a special source, e.g. spatially distributed matter, and then solve Einstein's field equation [\citenum{PN}-\citenum{ZR}]. Another method is to introduce quantum corrections to classical BHs [\citenum{AB}-\citenum{SB}]. BH solution with no essential singularity but with event horizon \cite{ANSOLDI} was first given by Bardeen \cite{BARDEEN}. Bardeen BH was later interpreted by Ay$o^{'}$n-Beato and Garc$i^{'}$a using field theory. We have observed significant progress in the study of non-rotating RBHs [\citenum{ID}-\citenum{ARUN}] as well as rotating RBHs [\citenum{ARUN}-\citenum{AE}].\\
BHs have strong gravitational field. It is because of this strong field, light rays close to BH are refracted and as a result, BH shadows are formed. The first image of a supermassive black hole M$87^*$ was unveiled by EHT \cite{KA, KA1}. Even before the first image given by EHT many attempts were made to study the observable appearance of a BH shadow. The shadow of a Kerr BH was studied by Bardeen et al. \cite{JM} whereas the shadow of a Schwarzschild black hole was examined by Synge \cite{JL}. Luminet studied BH surrounded by bright accretion disk \cite{JP}. The shadow radius is determined by the photon ring which is the characteristic of the underlying spacetime [\citenum{RN}]. Photon rings and shadow of RN black hole have been studied in \cite{YG}. Black hole shadows bear imprints of the underlying spacetime. Several studies have been conducted to detect dark matter using BH shadow [\citenum{RC}-\citenum{SN}].\\
Quasinormal modes are oscillations of a black hole that die down over time because of dissipative effects, e,g., emission of gravitational waves [\citenum{CV}-\citenum{CHANDRA}]. These are called quasinormal because of their transient nature. They are complex numbers where the real part signifies the frequency of the emitted gravitational wave and the imaginary part represents the decay rate or damping rate. Inspiral, merger, and ringdown are the phases that BHs experience after perturbation. For remnant BHs, quasinormal modes are equal to the ringdown phase. Quasinormal modes depend on the parameters of the BH. Thus, to study the underlying geometry it is important to investigate quasinormal modes. A significant number of studies have been conducted in this field [\citenum{CM}-\citenum{YY1}]. Hawking, by considering quantum consequences, showed that BHs emit radiation [\citenum{HAWKING}]. This radiation is known as Hawking radiation. When a pair production happens near the event horizon, one particle enters BH while the other moves away from BH. This second particle constitutes Hawking radiation [\citenum{HH}-\citenum{HC}]. Hawking temperature can be obtained through various methods [\citenum{SW}-\citenum{SI}]. The greybody factor is important in studying Hawking radiation. It can be calculated using the matching approach [\citenum{SF}-\citenum{JE}] or WKB approximation \cite{MK, CH}. An alternative to these methods was given by Visser \cite{GB} by finding rigorous bounds. The greybody factor was studied using this method in \cite{GB1, WJ}. \\
This manuscript is organized as follows. In section II, we introduce a non-rotating Simpson-Visser black hole and study the photonsphere and shadow of the black hole. Section III is devoted to studying quasinormal modes for scalar and electromagnetic perturbations and analyzing their graphical behavior. In section IV, we obtain expressions of the Hawking temperature and greybody bounds and study their variations. We conclude our article with conclusions in section V.

\section{Non-rotating Simpson-Visser black holes}
The concept of regular black holes was first proposed by Bardeen in his article \cite{BARDEEN}. Since then it has become a topic of great interest. One such regular black hole metric has been given by Simpson and Visser in their article \cite{SIMPSON}. The metric represents a non-rotating, static, and spherically symmetric black hole. The metric is defined by
\begin{equation}\label{SV}
ds^2=-(1-\frac{2M}{\sqrt{r^2+\alpha^2}})dt^2
+(1-\frac{2M}{\sqrt{r^2+\alpha^2}})^{-1}dr^2 + (r^2+\alpha^2)(d\theta^2+
sin^2\theta d\phi^2).
\end{equation}
Here, M is the ADM mass, and $\alpha$ is the parameter having a dimension of length. This solution encompasses both regular black holes as well as wormholes depending on the value of the parameter $\alpha$. We have a two-way, traversable wormhole
when $\alpha > 2M$ and a one-way wormhole with a null throat when
$\alpha = 2M$. The metric (\ref{SV}) gives a regular black hole when $\alpha<2M$. In this case, the singularity at $r=0$ is
replaced by a bounce to a different universe. The
bounce takes place through a spacelike throat shielded by an event
horizon and it is christened as black-bounce in \cite{SIMPSON}. The metric (\ref{SV}) reduces to that for the Schwarzschild black hole when we put $\alpha=0$.\\
In \cite{kirill}, authors have shown that we can obtain the metric (\ref{SV}) as an exact solution to Einstein's field equations minimally coupled with a self-interacting phantom scalar field $\varphi$ combined with a nonlinear electrodynamics field represented by tensor $F_{\mu\nu}$. The action is given by \cite{kirill}
\begin{equation}                       \label{Action}
S=\int\sqrt{-g}d^4r \Big(R + 2\ep g\MN \d_\mu\phi \d_\nu\varphi - 2V(\varphi) - \mathcal{L}(F) \Big),
\end{equation}
  where $\ep = \pm 1$. Here, $\mathcal{L}(F)$ is a gauge-invariant Lagrangian density and $F = F_{\mu\nu}F^{\mu\nu}$. We obtain the followin Einstein equation from the action (\ref{Action})
\begin{equation}                     \label{FieldEq}
               G\mN = - T\mN[\varphi] - T\mN[F],
\end{equation}  
  where stress-energy tensors, $T\mN[\varphi]$ and $T\mN[F]$, of the scalar and electromagnetic fields are given by
\begin{eqnarray}                 \label{SET-phi}\nonumber
			T\mN[\varphi]& =& 2\ep \d_{\mu}\varphi\d^{\nu}\varphi 
				- \delta\mN \left(\ep g^{\rho\sigma}\d_\rho \varphi \d_\sigma\varphi -V(\varphi)\right),\\\nonumber
			T\mN[F]& =& - 2 \frac{d\mathcal{L}}{dF} F_{\mu\sigma} F^{\nu\sigma} 
					+\frac 12 \delta\mN \mathcal{L}(F),
\end{eqnarray}
The expressions of $\varphi, V(\varphi),$ and $\mathcal{L}(F)$ are subsequently found and given in \cite{kirill}. It clealy shows that the metric (\ref{SV}) is an exact solution of Einstein equations with the action given by (\ref{Action}).\\
For the metric (\ref{SV}), the lapse function is given by
\begin{equation}
f(r)=1-\frac{2M}{\sqrt{r^2+\alpha^2}}.
\end{equation}
The nature of the lapse function is very important in calculating various observables related to the underlying spacetime. We graphically represent the variation of the lapse function with respect to r for various values of $\alpha$. From the plot, we observe that as we increase the value of $\alpha$, the position where the function crosses the r-axis shifts towards the left. It signifies a decrease in the position of the event horizon with an increase in the parameter value $\alpha$.
\begin{figure}[H]
\centering
\includegraphics[width=0.4\columnwidth]{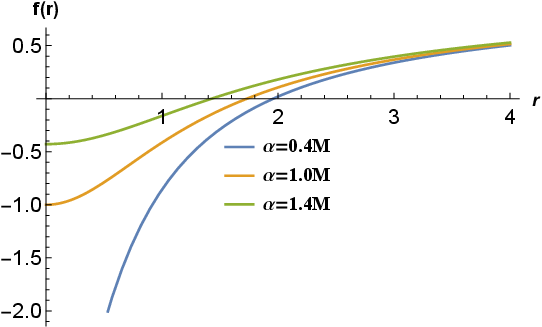}
\caption{Variation of lapse function with respect to r for various values of $\alpha$.}
\label{lfufig}
\end{figure}
The position of the event horizon is obtained by equating the lapse function to zero yields $r_h=\sqrt{4M^2-\alpha^2}$. It reinforces the finding from the above figure. Next, we move on to study null geodesics in the background of the spacetime given by ansatz (\ref{SV}). Since the spacetime we are considering is a spherically symmetric one, we can, without loss of generality, confine our study only to the equatorial plane. With this, the ansatz (\ref{SV}) reduces to
\begin{equation}\label{SV1}
ds^2=-f(r)dt^2+\frac{dr^2}{f(r)}+ h(r)d\phi^2,
\end{equation}
where $h(r)=r^2+\alpha^2$. The static and spherically symmetric nature of the spacetime ensures that the energy given by $\mathcal{E}=-p_{\mu} \xi_{(t)}^{\mu}$ and the angular momentum given by $\mathcal{L}=p_{\mu} \xi_{(\phi)}^{\mu}$ remain conserved along the geodesics. Here, $\xi_{(t)}^{\mu}$ and ${\xi_{(\phi)}^{\mu}}$ are the Killing vectors due to time-translational and rotational invariance respectively \cite{CHANDRA}. Thus, the energy of a photon is $\mathcal{E}=-p_t$, and the angular momentum is $\mathcal{L}=p_\phi$. To obtain the expressions of $p_t$ and $p_\phi$, we first write down the Lagrangian that corresponds to motion in the background of metric (\ref{SV1}). The Lagrangian is given by
\begin{equation}
\lagr=-f(r)\dot{t}^2+\frac{\dot{r}^2}{f(r)}+h(r)\dot{\phi}^2.
\end{equation}
With the help of definition $p_q=\frac{\partial \lagr}{\partial \dot{q}}$, we obtain
\begin{eqnarray}\nonumber
p_t&=&\frac{\partial \lagr}{\partial \dot{t}}=-f(r)\dot{t}, \\\nonumber
p_r&=&\frac{\partial \lagr}{\partial \dot{r}}=\frac{\dot{r}}{f(r)}, \\
p_\phi&=&\frac{\partial \lagr}{\partial \dot{\phi}}=h(r)\dot{\phi}.
\end{eqnarray}
Here, the dot is differentiation with respect to an affine parameter $\tau$. Thus, in terms of energy and angular momentum, we get two very important differential equations given by
\begin{equation}
\frac{d t}{d \tau}=\frac{\mathcal{E}}{f(r)} \quad \text{and} \quad \frac{d \phi}{d \tau}=\frac{\mathcal{L}}{h(r)}.\label{Conserved}
\end{equation}
Combining Eq.(\ref{Conserved}) and Eq.(\ref{SV1}), equation for the null geodesics is obtained as
\begin{equation}
\left(\frac{d r}{d \tau}\right)^{2} \equiv \dot{r}^{2}
=\mathcal{E}^{2}-V(r),
\end{equation}
where $V(r)$ is the effective potential given by
\begin{equation}
V(r)=\frac{\mathcal{L}^{2} f(r)}{h(r)}.
\end{equation}
The effective potential given above determines the motion of any particle in the underlying spacetime. We graphically show the variation of this potential.
\begin{figure}[H]
\centering
\includegraphics[width=0.4\columnwidth]{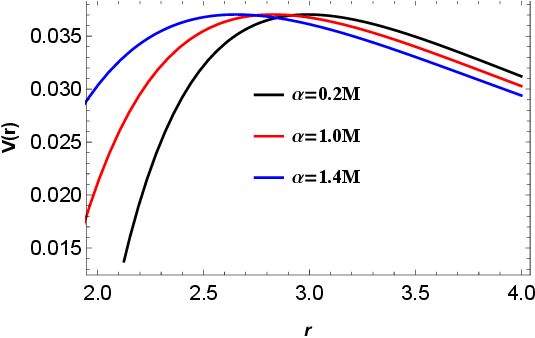}
\caption{Variation of potential with respect to r for various values of $\alpha$. Here, we have taken $\mathcal{L}=1$.}
\label{pot}
\end{figure}
The above plot indicates that the peak of the potential shifts towards the left as we increase the value of $\alpha$. This means the radius of the photonsphere decreases as we increase the value of $\alpha$.
For circular photon orbits of radius $r_{p}$, we must have
\begin{equation}
\frac{d V}{d r}|_{r=r_{p}}=0\Rightarrow \frac{f^{\prime}(r_p)}{f(r_p)}=\frac{h^{\prime}(r_p)}{h(r_p)}.
\end{equation}
Simple algebra produces $r_p=\sqrt{9M^2-b^2}$. This analytical expression confirms the inference we have drawn from Fig.(\ref{pot}). The corresponding impact parameter is
\begin{equation}
b_p=\frac{\mathcal{L}}{\mathcal{E}}=\sqrt{\frac{h(r_p)}{f(r_p)}}\Rightarrow b_p=3\sqrt{3},
\end{equation}
which is the same value as that for Schwarzschild black hole. Thus, we see that even though photon radius does depend on the parameter $\alpha$, the critical impact parameter does not depend on $\alpha$. Since, for a distant observer, the shadow radius is equal to the critical impact parameter, it is evident that the size of the shadow does not depend on the parameter $\alpha$.
\begin{figure}[H]
\centering
\includegraphics[width=0.4\columnwidth]{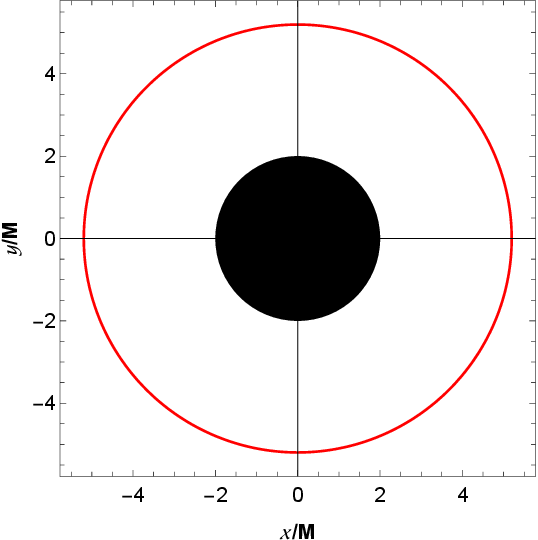}
\caption{Shadow of a non-rotating Simpson-Visser black hole.}
\label{shw}
\end{figure}
The region within the red dotted circle is the black hole shadow. The shadow cast by a black hole is larger than the actual size because of gravitational lensing.
\section{Quasinormal modes of non-rotating Simpson-Visser black hole}
In this section, we study quasinormal modes for scalar and electromagnetic perturbations of non-rotating Simpson-Visser black hole. Here, it is assumed that the impact of the scalar field or the electromagnetic on the background spacetime is negligible. To study quasinormal modes, we first consider the equation for the relevant field and then, reduce it to a Schr$\ddot{o}$dinger-like equation. For the scalar field, we will have the Klein-Gordon equation and for the electromagnetic field, we will consider Maxwell equations. For the massless scalar field, we have
\begin{eqnarray}
\frac{1}{\sqrt{-g}}{\partial \mu}(\sqrt{-g}g^{\mu\nu} \partial_{\nu}\chi) =0,
\label{scalar}
\end{eqnarray}
and for the electromagnetic field, we have
\begin{equation}
\frac{1}{\sqrt{-g}}{\partial \nu }(F_{\rho\sigma}g^{\rho\mu}g^{\sigma\nu}\sqrt{-g})=0,
\label{em}
\end{equation}
where $ F_{\rho\sigma}={\partial \rho}A^\sigma-{\partial \sigma}A^\rho $, $A_\nu$ being electromagnetic four-potential.
We now introduce the tortoise coordinate given by
\begin{eqnarray}
\text{d}r_{\ast}=\frac{\text{d}r}{f(r)}.
\label{tortoise}
\end{eqnarray}
We have $r_*\rightarrow -\infty$ as $r\rightarrow r_h$ and $r_*\rightarrow \infty$ as $r\rightarrow \infty$. With the help of tortoise coordinate, Eqs.(\ref{scalar}) and (\ref{em}) reduce to the Schr$\ddot{o}$dinger-like form given by
\begin{equation}
-\frac{\text{d}^2\phi}{\text{d}{r^2_*}}+V_{\text{eff}}(r) \phi=\omega ^{2}\phi,
\label{schrodinger}
\end{equation}
where the effective potential is given by
\begin{eqnarray}
V_{\text{eff}}(r)&=&\frac{(1-s^2)f(r)}{r}\frac{\text{d}f(r)}{\text{d}r}+\frac{f(r)\ell(\ell+1)}{r^{2}}\\\nonumber&=&\left(1-\frac{2M}{\sqrt{\alpha ^2+r^2}}\right) \left(\frac{\ell (\ell+1)}{r^2}+\frac{2 M\left(1-s^2\right)}{\left(\alpha ^2+r^2\right)^{3/2}}\right),
\label{vtotal}
\end{eqnarray}
where $\ell$ is the angular momentum and s is the spin. For $s=0$, we obtain the effective potential for scalar perturbation and for $s=1$, we obtain the effective potential for electromagnetic perturbation. Since the effective potential influences quasinormal modes, we briefly study the variation of the effective potential for various scenarios.
\begin{figure}[H]
\centering
\subfigure[]{
\label{vfig1}
\includegraphics[width=0.4\columnwidth]{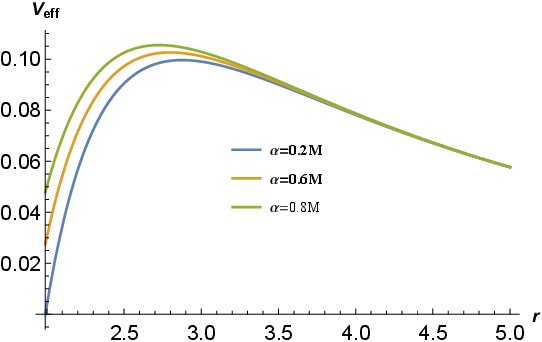}
}
\subfigure[]{
\label{vfig2}
\includegraphics[width=0.4\columnwidth]{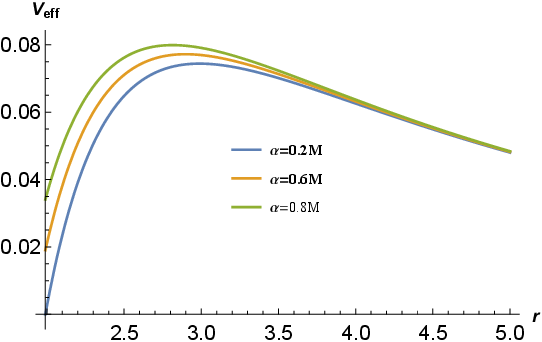}
}
\subfigure[]{
\label{vfig3}
\includegraphics[width=0.4\columnwidth]{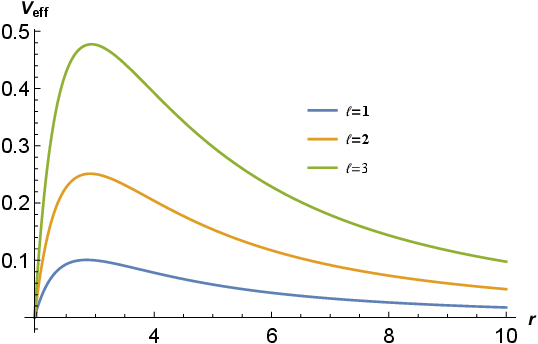}
}
\subfigure[]{
\label{vfig4}
\includegraphics[width=0.4\columnwidth]{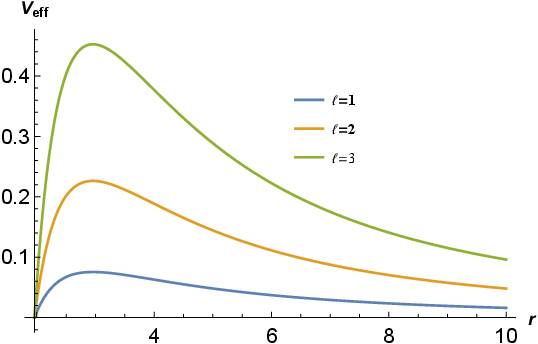}
}
\caption{Variation of effective potential with respect to normal coordinate r. The upper ones are for various values of $\alpha$ with $\ell=1$ and the lower ones are for various values of angular momentum with $\alpha=0.4M$. The left ones are for scalar perturbations and the right ones are for electromagnetic perturbations.}
\label{vfig}
\end{figure}
From above plots we observe that the peak of the effective potential increases with an increase in $\ell$ or $\alpha$. But the position of the peak shifts towards the right as we increase the angular momentum, whereas, for an increase in $\alpha$, the position shifts towards the left.

Schutz and Will in their article \cite{schutz} first developed the WKB method. Others extended the method to higher order WKB method \cite{iyer, iyer1, konoplya1}. For the 6th order WKB method, we have the expression of quasinormal frequencies as
\begin{equation}
\frac{\text{i}(\omega ^{2}-V_{0})}{\sqrt{-2V_{0}^{''}}}-\sum ^{6}_{\text{i}=2}\Omega_\text{i}=n+\frac{1}{2},
\label{WKB}
\end{equation}
where $V_{0}$ is the maximum value of the effective potential at the tortoise coordinate $r_\text{0}$, $V_\text{0}^{''}$ is the value of the second order derivative of the effective potential with respect to the tortoise coordinate evaluated at $r_\text{0}$, and $\Omega_\text{i}$ are the correction terms given in \cite{schutz, iyer, iyer1, konoplya1}. Now, to improve accuracy of WKB method, we employ Pad\'{e} approximants \cite{jerzy}, where in powers of the \emph{order parameter} $\epsilon$ we define a polynomial $P_k(\epsilon)$ as
\begin{equation}
P_k(\epsilon)=V_0+\Omega_2\epsilon^2+\Omega_4\epsilon^4+\Omega_6\epsilon^6+\ldots - i (n+\frac{1}{2})\sqrt{-2V_{0}^{''}}\left(\epsilon+\Omega_3\epsilon^3+\Omega_5\epsilon^5\ldots\right)
\end{equation}
Here, $k$ is the polynomial order, same as that for the WKB formula. We can obtain the squared frequency by putting $\epsilon =1$ via $\omega^2=P_{\tilde{n}/\tilde{m}}(1)$, where Pad\'{e} approximants, $P_{\tilde{n}/\tilde{m}}(\epsilon)$, for the polynomial $P_k(\epsilon)$, are given by \cite{jerzy, konoplya2}
\begin{equation}
    P_{\tilde{n}/\tilde{m}}(\epsilon)=\frac{\mathcal{Q}_0+\mathcal{Q}_1\epsilon+\ldots+\mathcal{Q}_{\tilde{n}}\epsilon^{\tilde{n}}}{\mathcal{R}_0+\mathcal{R}_1\epsilon+\ldots+\mathcal{R}_{\tilde{m}}\epsilon^{\tilde{m}}},
\end{equation}
with $\tilde{n}+\tilde{m}=k$. We use $6$th order Pad\'{e} averaged WKB approach to estimate the QNMs and tabulate some of the values of quasinormal frequencies of scalar and electromagnetic perturbations for various values of angular momentum $\ell$ and parameter $\alpha$. Here, we take $M=1$ for all calculations. In Table \ref{QNMS}, we show quasinormal modes of scalar perturbation for different values of angular momentum $\ell$ and parameter $\alpha$ keeping overtone number $n=0$. iIn Table \ref{QNMSN}, we show those for $n=1$. In Table \ref{QNMEM}, we show quasinormal modes of electromagnetic perturbation for different values of angular momentum and parameter $\alpha$ keeping overtone number $n=0$ and in Table \ref{QNMEMN}, we show those for $n=1$.
\begin{table}[!htp]
\centering
\caption{Quasinormal frequencies for scalar field with $n=0$.}
\setlength{\tabcolsep}{3mm}
\begin{tabular}{|c|c|c|c|}
\hline
\text{$\alpha/M $ } & \text{$\ell $=1} & \text{$\ell $=2} & \text{$\ell $=3} \\
\hline
 0. & 0.292931\, -0.097661 i & 0.483644\, -0.0967591 i & 0.675366\, -0.0964997 i \\
\hline
 0.2 & 0.293625\, -0.0975045 i & 0.484747\, -0.0966098 i & 0.676891\, -0.0963529 i \\
\hline
 0.4 & 0.295746\, -0.0970231 i & 0.488122\, -0.0961494 i & 0.681554\, -0.0959 i \\
\hline
 0.6 & 0.299405\, -0.0961771 i & 0.493966\, -0.0953383 i & 0.689639\, -0.0951003 i \\
\hline
 0.8 & 0.304815\, -0.0948936 i & 0.502661\, -0.0940996 i & 0.701683\, -0.0938749 i \\
\hline
 1. & 0.312351\, -0.0930422 i & 0.514855\, -0.0922961 i & 0.71861\, -0.0920829 i \\
\hline
\end{tabular}
\label{QNMS}
\end{table}

\begin{table}[!htp]
\centering
\caption{Quasinormal frequencies for scalar field with $n=1$.}
\setlength{\tabcolsep}{3mm}
\begin{tabular}{|c|c|c|c|}
\hline
\text{$\alpha/M $ } & \text{$\ell $=1} & \text{$\ell $=2} & \text{$\ell $=3} \\
\hline
 0. & 0.264456\, -0.306507 i & 0.463846\, -0.295626 i & 0.660671\, -0.292288 i \\
\hline
 0.2 & 0.265441\, -0.305916 i & 0.465144\, -0.295125 i & 0.662339\, -0.291819 i \\
\hline
 0.4 & 0.268413\, -0.304082 i & 0.469099\, -0.293584 i & 0.667432\, -0.290377 i \\
\hline
 0.6 & 0.273452\, -0.300813 i & 0.475912\, -0.290882 i & 0.676236\, -0.287835 i \\
\hline
 0.8 & 0.280786\, -0.295861 i & 0.485958\, -0.286764 i & 0.68929\, -0.283953 i \\
\hline
 1. & 0.290766\, -0.288748 i & 0.499899\, -0.280815 i & 0.707525\, -0.278302 i \\
\hline
\end{tabular}
\label{QNMSN}
\end{table}

\begin{table}[!htp]
\centering
\caption{Quasinormal frequencies for electromagnetic field with $n=0$.}
\setlength{\tabcolsep}{3mm}
\begin{tabular}{|c|c|c|c|}
\hline
\text{$\alpha/M $ } & \text{$\ell $=1} & \text{$\ell $=2} & \text{$\ell $=3} \\
\hline
 0. & 0.248251\, -0.0924847 i & 0.457595\, -0.0950048 i & 0.656899\, -0.0956163 i \\
\hline
 0.2 & 0.249003\, -0.0923694 i & 0.458724\, -0.0948656 i & 0.65844\, -0.0954741 i \\
\hline
 0.4 & 0.251304\, -0.0920112 i & 0.462177\, -0.0944359 i & 0.663153\, -0.095035 i \\
\hline
 0.6 & 0.255271\, -0.0913617 i & 0.468163\, -0.0936765 i & 0.671328\, -0.0942589 i \\
\hline
 0.8 & 0.261148\, -0.0903477 i & 0.477077\, -0.0925117 i & 0.683516\, -0.0930677 i \\
\hline
 1. & 0.269347\, -0.0888389 i & 0.489602\, -0.0908062 i & 0.700663\, -0.0913222 i \\
\hline
\end{tabular}
\label{QNMEM}
\end{table}

\begin{table}[!htp]
\centering
\caption{Quasinormal frequencies for electromagnetic field with $n=1$.}
\setlength{\tabcolsep}{3mm}
\begin{tabular}{|c|c|c|c|}
\hline
\text{$\alpha/M $ } & \text{$\ell $=1} & \text{$\ell $=2} & \text{$\ell $=3} \\
\hline
 0. & 0.214272\, -0.294104 i & 0.436533\, -0.290727 i & 0.641736\, -0.289731 i \\
\hline
 0.2 & 0.215404\, -0.293678 i & 0.437875\, -0.290256 i & 0.643427\, -0.289276 i \\
\hline
 0.4 & 0.21872\, -0.292259 i & 0.441961\, -0.288802 i & 0.648591\, -0.287872 i \\
\hline
 0.6 & 0.224396\, -0.289778 i & 0.448999\, -0.286244 i & 0.657517\, -0.285398 i \\
\hline
 0.8 & 0.2325\, -0.28546 i & 0.459381\, -0.282324 i & 0.67076\, -0.281612 i \\
\hline
 1. & 0.243736\, -0.279086 i & 0.4738\, -0.276634 i & 0.689272\, -0.276092 i \\
\hline
\end{tabular}
\label{QNMEMN}
\end{table}
From above tables, we can infer that the real part of quasinormal frequencies increases with an increase in parameter value $\alpha$ for a particular value of $\ell$. Additionally, it is observed for both perturbations that the real part of quasinormal modes increases as we increase the angular momentum $\ell$. We can observe from the Table (\ref{QNMS}) and Table (\ref{QNMSN}) that the decay rate or damping rate increases as we decrease the value of parameter $\alpha$ or the angular momentum for scalar perturbation. From Tables (\ref{QNMEM}) and (\ref{QNMEMN}) we can infer that the damping rate or decay rate increases as we decrease the value of the parameter $\alpha$ or increase the angular momentum for electromagnetic perturbation. If we compare values of quasinormal modes for different overtone numbers, then we can see that the real part of quasinormal modes decreases with the overtone number but the decay or damping rate increases with the overtone number. \\
Next, to understand the variation of error associated with an order WKB method, we tabulate quasinormal frequency for various orders of the WKB method and corresponding error associated with the method. The error is measured by the formula $\text{error}=\frac{|\omega_{k+1}-\omega_{k-1}|}{2}$. Here, we have considered scalar perturbation with $n=0$, $\alpha=0.2M$, and $\ell=1$.
\begin{table}[!htp]
\centering
\caption{Quasinormal frequency and error for various orders of WKB approximation.}
\setlength{\tabcolsep}{3mm}
\begin{tabular}{|c|c|c|c|}
\hline
\text{WKB Order } & \text{Quasinormal frequency} & \text{Error} \\
\hline
8 & 0.294678\, -0.0971196 i & 0.030877 \\
\hline
7 & 0.293563\, -0.0974886 i & 0.000590874 \\
\hline
6 & 0.293603\, -0.097611 i & 0.000108075 \\
\hline
5 & 0.293768\, -0.0975564 i & 0.000199927 \\
\hline
4 & 0.293655\, -0.0972145 i & 0.000987336 \\
\hline
3 & 0.291812\, -0.0978283 i & 0.00516503 \\
\hline
\end{tabular}
\label{QNMER}
\end{table}
We can observe from the above table that the error decreases as we increase the order of WKB approximation upto sixth order and then, it starts increasing. Thus, we can infer that sixth order WKB approximation gives the best value of quasinormal frequency. Now, we graphically show the variation of quasinormal frequency for various aspects.
\begin{figure}[H]
\centering
\subfigure[]{
\label{qnmimfig1}
\includegraphics[width=0.4\columnwidth]{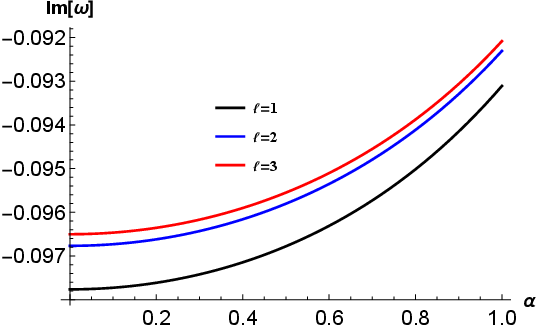}
}
\subfigure[]{
\label{qnmimfig2}
\includegraphics[width=0.4\columnwidth]{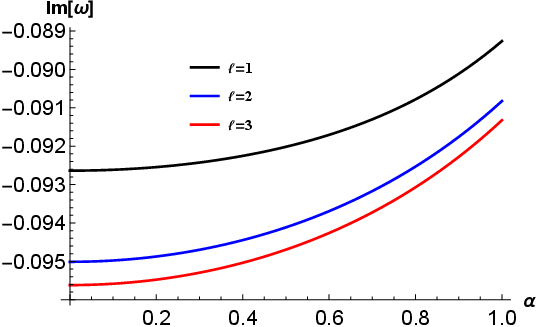}
}
\caption{It gives the variation of the imaginary part of the quasinormal frequency with respect to $\alpha$ for various values of $\ell$. The left one is for the scalar field and the right one is for the electromagnetic field.}
\label{qnmimfig}
\end{figure}

\begin{figure}[H]
\centering
\subfigure[]{
\label{qnmrefig1}
\includegraphics[width=0.4\columnwidth]{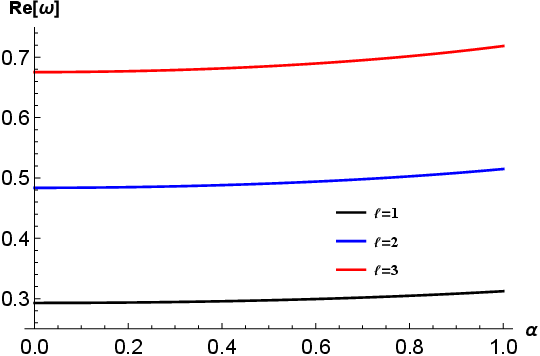}
}
\subfigure[]{
\label{qnmrefig2}
\includegraphics[width=0.4\columnwidth]{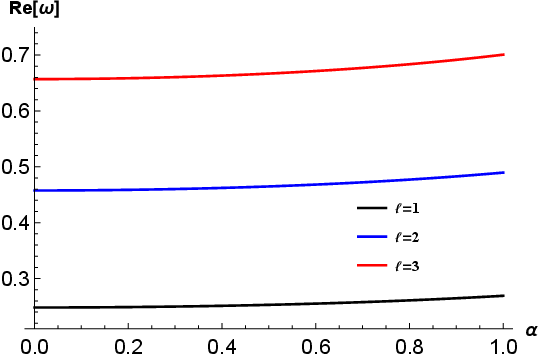}
}
\caption{It gives the variation of the real part of the quasinormal frequency with respect to $\alpha$ for various values of $\ell$. The left one is for the scalar field and the right one is for the electromagnetic field.}
\label{qnmrefig}
\end{figure}

\begin{figure}[H]
\centering
\subfigure[]{
\label{qnmrefig1}
\includegraphics[width=0.4\columnwidth]{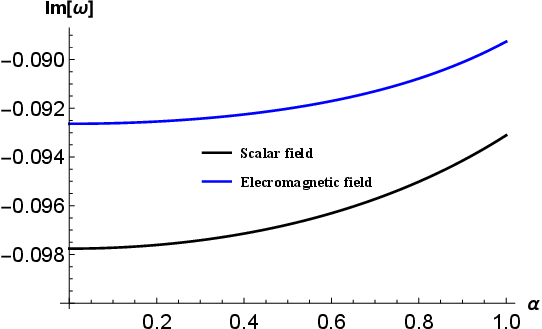}
}
\subfigure[]{
\label{qnmrefig2}
\includegraphics[width=0.4\columnwidth]{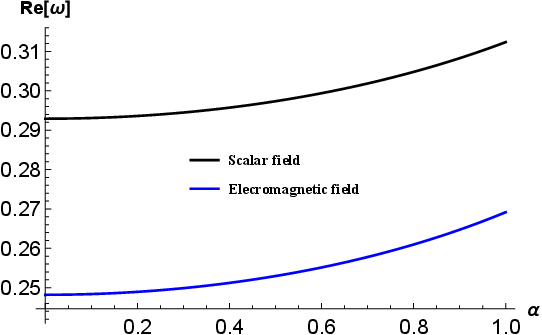}
}
\caption{Left one gives the variation of the imaginary part of the quasinormal frequency with respect to $\alpha$ for scalar and electromagnetic fields and the right one gives that for the real part. Here, we have taken $\ell=1$.}
\label{qnmimrefig}
\end{figure}
Fig.(\ref{qnmimfig}) and Fig.(\ref{qnmrefig}) reinforce findings we have drwan from Tabs.(\ref{QNMS}, \ref{QNMSN}, \ref{QNMEM}, \ref{QNMEMN}). We can also observe that the real part of quasinormal modes is larger for scalar perturbation, whereas, the imaginary part is larger for electromagnetic perturbation. It implies that the damping rate or decay rate is larger for scalar perturbation. We next study the convergence of the WKB method for various values of $(n,\ell)$ pair.

\begin{figure}[H]
\centering
\subfigure[]{
\label{qnmorderfig1}
\includegraphics[width=0.4\columnwidth]{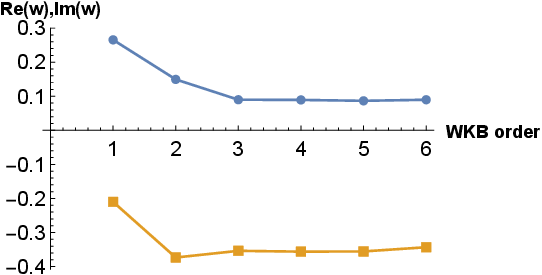}
}
\subfigure[]{
\label{qnmorderfig2}
\includegraphics[width=0.4\columnwidth]{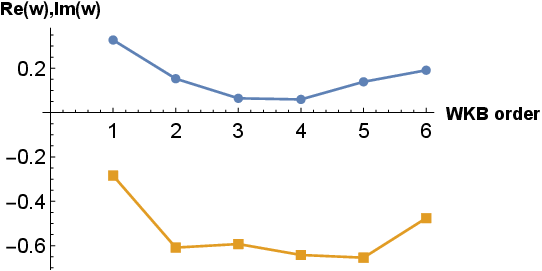}
}
\subfigure[]{
\label{qnmorderfig3}
\includegraphics[width=0.4\columnwidth]{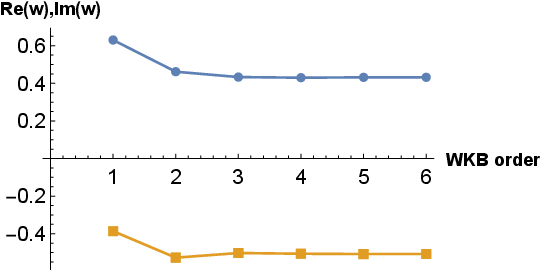}
}
\subfigure[]{
\label{qnmorderfig4}
\includegraphics[width=0.4\columnwidth]{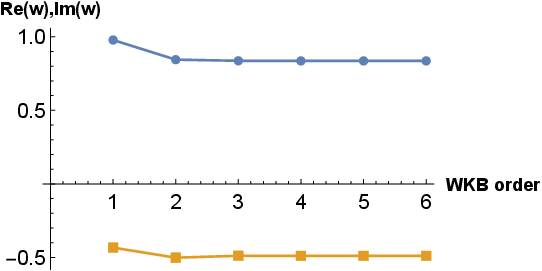}
}
\caption{Variation of the real and imaginary part of quasinormal frequencies with respect to WKB order for various values of $(n,\ell)$ pair. The upper left one is for $(1,0)$ pair, the upper right one is for $(2,0)$ pair, the lower left one is for $(2,2)$ pair and the lower right one is for $(2,4)$ pair. The upper line in each plot is for the real part of quasinormal modes and the lower line is for the imaginary part of quasinormal modes.}
\label{qnmorderfig}
\end{figure}
From the above figure we observe that quasinormal frequencies fluctuate even for higher order when we consider the pair $(2,0)$. This confirms the finding in the article \cite{KD} where it is observed that WKB approximation is reliable when the angular momentum is high and the overtone number is low.
\section{Hawking temperature and Bounds of the Greybody factor}
In this section, we intend to calculate the Hawking temperature and greybody bounds for the black hole under consideration. Hawking in his article \cite{HAWKING} showed that black holes emit radiation. That radiation is known as Hawking radiation. Bekenstein in his article \cite{BEK} and Keifer in his article \cite{KEIF} showed that it was necessary to associate a temperature with the horizon for consistency with thermodynamics. The Hawking temperature is given by
\begin{equation}
T_H=\frac{1}{4\pi \sqrt{-g_{tt}g_{rr}}}\frac{dg_{tt}}{dr}|_{r=r_h}.
\end{equation}
For the metric in consideration, we have $g_{tt}=-f(r)$ and $g_{rr}=\frac{1}{f(r)}$. Putting these values in the above equation, we get
\begin{equation}
T_H=\frac{\sqrt{4 M^2-\alpha ^2}}{16 \pi M^2}.
\end{equation}
The dependence of the Hawking temperature on the parameter $\alpha$ is evident from the above equation. We recover the value of the Hawking temperature for the Schwarzschild black hole if we put $\alpha=0$ in the above equation. To show the dependence graphically, we plot the Hawking temperature against $\alpha$.
\begin{figure}[H]
\centering
\subfigure[]{
\label{hw1}
\includegraphics[width=0.4\columnwidth]{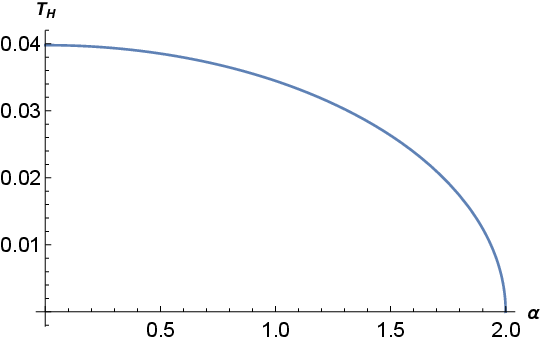}
}
\caption{Variation of Hawking temperature with respect to $\alpha$.}
\label{hwfig}
\end{figure}
We can observe that the Hawking temperature decreases as we increase the value of the parameter $\alpha$. The Hawking radiation observed by an asymptotic observer is different from the original radiation near the horizon of the black hole due to the redshift factor. Greybody distribution describes the Hawking radiation that is observed by an asymptotic observer. Here, we try to obtain the lower bound of the greybody factor for a non-rotating Simpson-Visser black hole. A lot of research has been dedicated to the bound greybody factor. Visser and Boonserm in their articles \cite{GB, GB1, GB2} gave an elegant way to lower bound the greybody factor. A rigorous bound of the transmission probability, which is the same as that of the graybody factor, is given by
\begin{equation}
T\geq sech^{2}(\frac{1}{2\omega}\int_{-\infty}^{\infty}|V_{\text{eff}}(r_*)|dr_*),\label{grey}
\end{equation}
where $r_*$ is the tortoise coordinate defined in Eq.(\ref{tortoise}) and $V_{\text{eff}}(r_*)$ is the potential given in Eq.(\ref{vtotal}). In terms of normal coordinate r, the above equation becomes
\begin{equation}
T\geq sech^{2}(\frac{1}{2\omega}\int_{r_h}^{\infty}|V_{\text{eff}}(r)|\frac{dr}{f(r)})\label{grey1}.
\end{equation}
If we use Eq.(\ref{vtotal}), then, the above equation reduces to
\begin{equation}
T\geq sech^2\left(\frac{\frac{\ell (\ell+1)}{\sqrt{4-\alpha^2}}+\frac{1-s^2}{\sqrt{4-\alpha^2}+2}}{2 \omega }\right).
\end{equation}
The above equation shows the explicit dependence of greybody factor bounds on the value of parameter $\alpha$. We, next, plot the bounds of the greybody factor against $\omega$ for both scalar and electromagnetic perturbations by taking $s=0$ and $s=1$ respectively. We have plotted T for various values of $\alpha$ as well as angular momentum $\ell$.
\begin{figure}[H]
\centering
\subfigure[]{
\label{gbsfig1}
\includegraphics[width=0.4\columnwidth]{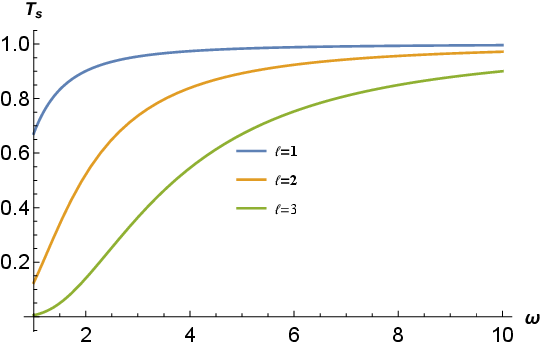}
}
\subfigure[]{
\label{gbemfig1}
\includegraphics[width=0.4\columnwidth]{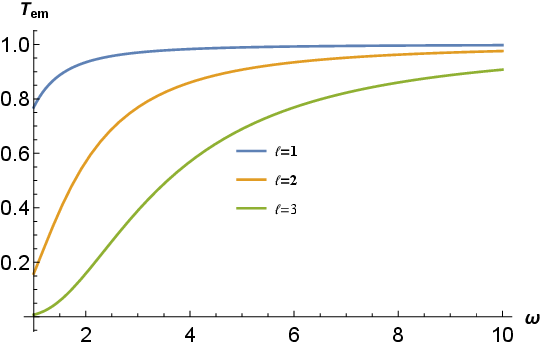}
}
\caption{Bounds of greybody factor for various values of $\ell$. Left one is for scalar perturbation and the right one is for electromagnetic perturbation. Here we have taken $\alpha=0.6$.}
\label{gbfig}
\end{figure}

\begin{figure}[H]
\centering
\subfigure[]{
\label{gbsfig2}
\includegraphics[width=0.4\columnwidth]{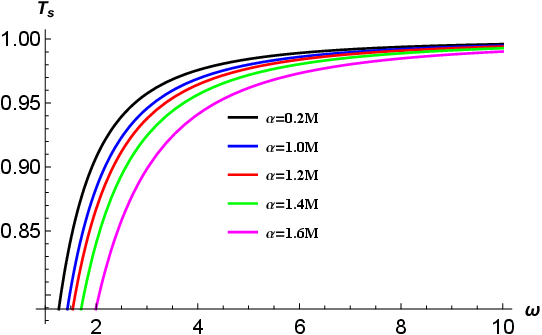}
}
\subfigure[]{
\label{gbsfig2}
\includegraphics[width=0.4\columnwidth]{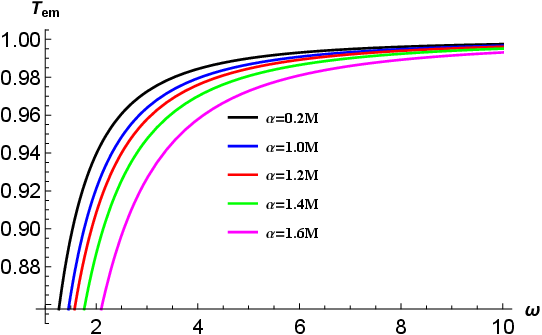}
}
\caption{Bounds of greybody factor for various values of $\alpha$. Left one is for scalar perturbation and the right one is for electromagnetic perturbation. Here we have taken $\ell=1$.}
\label{gbfig1}
\end{figure}
Here, $T_s$ is the greybody factor for scalar perturbation, and $T_{em}$ is the greybody factor for electromagnetic perturbation. There are a few conclusions we can draw from the above plots. We can observe that the bounds asymptotically approach the value $1$, for both scalar and electromagnetic perturbations. The nature of variation is also the same for both perturbations. From Fig.(\ref{gbfig}) we can conclude that the bounds decrease as we increase the angular momentum. We can also conclude from Fig.(\ref{gbfig1}) that the bounds decrease as we increase the value of the parameter $\alpha$.
\section{Conclusions}
We have conducted an investigation of photonsphere, shadow radius, quasinormal modes, Hawking temperature, and greybody bounds of non-rotating Simpson-Visser black hole. To obtain the radius of the photosphere, we first write down the Lagrangian corresponding to the metric confining ourselves to the equatorial plane without loss of generality. Then, with the help of Killing vectors and the Lagrangian, we obtain the differential equation of motion for photons and get the expression of the effective potential. Imposing conditions on the effective potential and its derivative for circular photon orbits, we obtain the expression of the radius of photonsphere $r_p$ and the corresponding impact parameter. For an observer located at asymptotic infinity, the shadow radius is equal to the impact parameter. We observe that the radius of the photonsphere decreases with an increase in the parameter $\alpha$, whereas, the shadow radius remains constant to the value $3\sqrt{3}$ which is the shadow radius for Schwarzschild black hole. \\
Next, we study quasinormal modes for two types of perturbations: scalar and electromagnetic. We tabulate quasinormal frequencies for various values of overtone number n, angular momentum $\ell$, and parameter $\alpha$ in Tables (\ref{QNMS}, \ref{QNMSN}, \ref{QNMEM}, \ref{QNMEMN}, \ref{QNMER}). Our findings indicate that with the increase in the value of $\alpha$, the real part of quasinormal frequency increases. We observe that for scalar and electromagnetic perturbations, the decay rate or damping rate increases as we decrease the value of parameter $\alpha$. But, the decay rate or the damping rate increases for scalar perturbation and decreases for electromagnetic perturbation as we decrease the angular momentum. Comparing values of quasinormal modes for different overtone numbers, we find that the real part of quasinormal modes decreases with the overtone number but the decay or damping rate increases with the overtone number. These findings are reinforced in various plots presented. We also observe that the error associated with a WKB order decreases as we increase the order of WKB approximation upto the sixth order and then it starts increasing. Moreover, the real part of quasinormal modes as well as the decay rate or damping rate for scalar perturbation is greater than that of electromagnetic perturbation. Our findings also confirm the result in the article \cite{KD} where it was observed that WKB approximation is reliable when the angular momentum is high and the overtone number is low.
Finally, we have obtained the expression of Hawking temperature and bounds of the greybody factor. Our findings indicate that the Hawking temperature decreases with an increase in the parameter $\alpha$. On the other hand, the bounds of the greybody factor for both type of perturbations decreases with an increase in angular momentum $\ell$ or parameter $\alpha$. We can obtain further insights into the behavior of various perturbations with the help of time-domain calculations and innovative approaches. We also hope that results from future experiments will guide us to have a complete theory of quantum gravity.\\
\\
\textbf{Data Availability Statement}: We do not have any additional data to present


\begin{thebibliography}{the}
\bibitem{EINSTEIN}Albert Einstein, “Lens-Like Action of a Star by the Deviation of Light in the Gravitational Field,” Science 84, 506–507 (1936).
\bibitem{PN}P. Nicolini, A. Smailagic, and E. Spallucci, “Noncommutative geometry inspired
Schwarzschild black hole,” Phys. Lett. B 632 (2006) 547–551, arXiv:gr-qc/0510112.
\bibitem{PN1} P. Nicolini, “Noncommutative Black Holes, The Final Appeal To Quantum Gravity: A
Review,” Int. J. Mod. Phys. A 24 (2009) 1229–1308, arXiv:0807.1939 [hep-th].
\bibitem{EP} E. Spallucci, A. Smailagic, and P. Nicolini, “Non-commutative geometry inspired
higher-dimensional charged black holes,” Phys. Lett. B 670 (2009) 449–454,
arXiv:0801.3519 [hep-th].
\bibitem{PN2} P. Nicolini and E. Spallucci, “Noncommutative geometry inspired wormholes and dirty
black holes,” Class. Quant. Grav. 27 (2010) 015010, arXiv:0902.4654 [gr-qc].
\bibitem{AB} A. B. Balakin and A. E. Zayats, “Non-minimal Wu-Yang monopole,” Phys. Lett. B 644
(2007) 294–298, arXiv:gr-qc/0612019.
\bibitem{AB1} A. B. Balakin, J. P. S. Lemos, and A. E. Zayats, “Magnetic black holes and monopoles in
a nonminimal Einstein-Yang-Mills theory with a cosmological constant: Exact solutions,”
Phys. Rev. D 93 no. 8, (2016) 084004, arXiv:1603.02676 [gr-qc].
\bibitem{ZR} Z. Roupas, “Detectable universes inside regular black holes,” Eur. Phys. J. C 82 no. 3,
(2022) 255, arXiv:2203.13295 [gr-qc].
\bibitem{AB2} A. Bonanno and M. Reuter, “Renormalization group improved black hole space-times,”
Phys. Rev. D 62 (2000) 043008, arXiv:hep-th/0002196.
30
\bibitem{LM} L. Modesto, “Disappearance of black hole singularity in quantum gravity,” Phys. Rev. D
70 (2004) 124009, arXiv:gr-qc/0407097.
\bibitem{RG} R. Gambini and J. Pullin, “Black holes in loop quantum gravity: The Complete
space-time,” Phys. Rev. Lett. 101 (2008) 161301, arXiv:0805.1187 [gr-qc].
\bibitem{BK} B. Koch and F. Saueressig, “Black holes within Asymptotic Safety,” Int. J. Mod. Phys. A
29 no. 8, (2014) 1430011, arXiv:1401.4452 [hep-th].
\bibitem{AP} A. Perez, “Black Holes in Loop Quantum Gravity,” Rept. Prog. Phys. 80 no. 12, (2017)
126901, arXiv:1703.09149 [gr-qc].
\bibitem{NB} N. Bodendorfer, F. M. Mele, and J. M¨unch, “Mass and Horizon Dirac Observables in
Effective Models of Quantum Black-to-White Hole Transition,” Class. Quant. Grav. 38
no. 9, (2021) 095002, arXiv:1912.00774 [gr-qc].
\bibitem{NB1} N. Bodendorfer, F. M. Mele, and J. M¨unch, “(b,v)-type variables for black to white hole
transitions in effective loop quantum gravity,” Phys. Lett. B 819 (2021) 136390,
arXiv:1911.12646 [gr-qc].
\bibitem{MB} M. Bojowald, “Black-Hole Models in Loop Quantum Gravity,” Universe 6 no. 8, (2020)
125, arXiv:2009.13565 [gr-qc].
\bibitem{SB} S. Brahma, C.-Y. Chen, and D.-h. Yeom, “Testing Loop Quantum Gravity from
Observational Consequences of Nonsingular Rotating Black Holes,” Phys. Rev. Lett. 126
no. 18, (2021) 181301, arXiv:2012.08785 [gr-qc].
\bibitem{PENROSE} Roger Penrose, “Gravitational collapse and space-time singularities,” Phys. Rev. Lett. 14, 57–59 (1965).
\bibitem{ANSOLDI} Stefano Ansoldi, “Spherical black holes with regular center: A Review of existing models including a recent realization with Gaussian
sources,” in Conference on Black Holes and Naked Singularities (2008) arXiv:0802.0330 [gr-qc].
\bibitem{BARDEEN} J. M. Bardeen, “Non-singular general-relativistic gravitational collapse,” Proceedings of GR5 URSS, Tbilisi , 174 (1968).
\bibitem{ELOY} Eloy Ayon-Beato and Alberto Garca, “New regular black hole solution from nonlinear electrodynamics.” Phys. Lett. B 464, 25–29 (1999).
\bibitem{ID} I. Dymnikova, “Vacuum nonsingular black hole.” Gen Relat Gravit 24, 235242 (1992).
\bibitem{JOS} Jos´e P. S. Lemos and Vilson T. Zanchin, “Regular black holes: Electrically charged solutions, reissner-nordstr¨om outside a de sitter core,”
Phys. Rev. D 83, 124005 (2011).
\bibitem{ARUN} Arun Kumar, Sushant G. Ghosh, and Sunil D. Maharaj, “Nonsingular black hole chemistry,” Physics of the Dark Universe 30, 100634
(2020).
\bibitem{AE} A. Eichhorn and A. Held, “Image features of spinning regular black holes based on a locality principle.” Eur. Phys. J. C 81, 933 (2021).
\bibitem{KA}Kazunori Akiyama et al. (Event Horizon Telescope), “First M87 Event Horizon Telescope Results. I. The Shadow of the Supermassive
Black Hole,” Astrophys. J. Lett. 875, L1,17 (2019).
\bibitem{KA1} Kazunori Akiyama et al. (Event Horizon Telescope), “First Sagittarius A* Event Horizon Telescope Results. I. The Shadow of the
Supermassive Black Hole in the Center of the Milky Way,” Astrophys. J. Lett. 930, L12 (2022).
\bibitem{JM} James M. Bardeen, William H. Press, and Saul A. Teukolsky, “Rotating Black Holes: Locally Nonrotating Frames, Energy Extraction,
and Scalar Synchrotron Radiation,” Astrophys. J. 178, 347–370 (1972).
\bibitem{JL} J. L. Synge, “The Escape of Photons from Gravitationally Intense Stars,” Mon. Not. Roy. Astron. Soc. 131, 463–466 (1966).
\bibitem{JP} J. P. Luminet, “Image of a spherical black hole with thin accretion disk,” Astron. Astrophys. 75, 228–235 (1979).
\bibitem{RN} Ramesh Narayan, Michael D. Johnson, and Charles F. Gammie, “The shadow of a spherically accreting black hole,” The Astrophysical
Journal 885, L33 (2019).
\bibitem{YG} Yang Guo and Yan-Gang Miao, “Charged black-bounce spacetimes: Photon rings, shadows and observational appearances,” Nucl. Phys. B
983, 115938 (2022), arXiv:2112.01747 [gr-qc].
\bibitem{RC} Reggie C. Pantig, Paul K. Yu, Emmanuel T. Rodulfo, and Ali O¨ vgu¨n, “Shadow and weak deflection angle of extended uncertainty principle
black hole surrounded with dark matter,” Annals Phys. 436, 168722 (2022), arXiv:2104.04304 [gr-qc].
\bibitem{RA} R. A. Konoplya and A. Zhidenko, “Solutions of the Einstein Equations for a Black Hole Surrounded by a Galactic Halo,” Astrophys. J.
933, 166 (2022), arXiv:2202.02205 [gr-qc].
\bibitem{RA1} R. A. Konoplya, “Shadow of a black hole surrounded by dark matter,” Phys. Lett. B 795, 1–6 (2019), arXiv:1905.00064 [gr-qc].
\bibitem{ZX} Zhaoyi Xu, Xian Hou, Xiaobo Gong, and Jiancheng Wang, “Black Hole Space-time In Dark Matter Halo,” JCAP 09, 038 (2018).
\bibitem{ZX1} Zhaoyi Xu, Xiaobo Gong, and Shuang-Nan Zhang, “Black hole immersed dark matter halo,” Phys. Rev. D 101, 024029 (2020).
\bibitem{RC1} Reggie C. Pantig and Emmanuel T. Rodulfo, “Rotating dirty black hole and its shadow,” Chin. J. Phys. 68, 236–257 (2020),
arXiv:2003.06829 [gr-qc].
\bibitem{WJ} Wajiha Javed, Hafsa Irshad, Reggie C. Pantig, and Ali vgn, “Weak deflection angle by kalb-ramond traversable wormhole in plasma and
dark matter mediums,” Universe 8 (2022), 10.3390/universe8110599.
\bibitem{WJ1} Wajiha Javed, Sibgha Riaz, Reggie C. Pantig, and Ali O¨ vgu¨n, “Weak gravitational lensing in dark matter and plasma mediums for
wormhole-like static aether solution,” Eur. Phys. J. C 82, 1057 (2022), arXiv:2212.00804 [gr-qc].
\bibitem{KJ} Kimet Jusufi, Mubasher Jamil, and Tao Zhu, “Shadows of Sgr $A^*$ black hole surrounded by superfluid dark matter halo,” Eur. Phys. J. C
80, 354 (2020), arXiv:2005.05299 [gr-qc].
\bibitem{SN} Sourabh Nampalliwar, Saurabh Kumar, Kimet Jusufi, Qiang Wu, Mubasher Jamil, and Paolo Salucci, “Modeling the Sgr A* Black Hole
Immersed in a Dark Matter Spike,” Astrophys. J. 916, 116 (2021), arXiv:2103.12439 [astro-ph.HE].
\bibitem{CV}C. V. Vishveshwara, Stability of the Schwarzschild Metric, Phys. Rev. D 1, 2870 (1970).
\bibitem{WH} W. H. Press, Long Wave Trains of Gravitational Waves from a Vibrating Black Hole, ApJ 170, L105 (1971).
\bibitem{CHANDRA} S. Chandrasekhar and S. Detweiler, The Quasi-Normal Modes of the Schwarzschild Black Hole, Proc. R. Soc. Lond. A 344, 441 (1975).
\bibitem{CM} C. Ma, Y. Gui, W. Wang, F. Wang, Massive scalar field quasinormal modes of a Schwarzschild black hole surrounded by quintessence,
Cent. Eur. J. Phys. 6, 194 (2008) [arXiv:gr-qc/0611146].
\bibitem{DJ} D. J. Gogoi and U. D. Goswami, A New f(R) Gravity Model and Properties of Gravitational Waves in It, Eur. Phys. J. C 80, 1101 (2020)
[arXiv:2006.04011].
\bibitem{DJ1} D. J. Gogoi and U. D. Goswami, Gravitational Waves in f (R) Gravity Power Law Model, Indian J. Phys. 96, 637 (2022)
[arXiv:1901.11277].
\bibitem{DL} D. Liang, Y. Gong, S. Hou and Y. Liu, Polarizations of Gravitational Waves in f(R) Gravity, Phys. Rev. D 95, 104034 (2017)
[arXiv:1701.05998].
\bibitem{RO} R. Oliveira, D. M. Dantas, and C. A. S. Almeida, Quasinormal Frequencies for a Black Hole in a Bumblebee Gravity, EPL 135, 10003
(2021) [arXiv:2105.07956].
\bibitem{DJ2} D. J. Gogoi and U. D. Goswami, Quasinormal Modes of Black Holes with Non-Linear-Electrodynamic Sources in Rastall Gravity, Physics
of the Dark Universe 33, 100860 (2021) [arXiv:2104.13115].
\bibitem{JP1} J. P. M. Grac¸a and I. P. Lobo, Scalar QNMs for Higher Dimensional Black Holes Surrounded by Quintessence in Rastall Gravity, Eur.
Phys. J. C 78, 101 (2018) [arXiv:1711.08714].
\bibitem{YZ} Y. Zhang, Y.X. Gui, F. Li, Quasinormal modes of a Schwarzschild black hole surrounded by quintessence: electromagnetic perturbations,
Gen. Relativ. Gravit. 39, 1003 (2007) [arXiv:gr-qc/0612010].
\bibitem{MB1} M. Bouhmadi-L´opez, S. Brahma, C.-Y. Chen, P. Chen, and D. Yeom, A Consistent Model of Non-Singular Schwarzschild Black Hole in
Loop Quantum Gravity and Its Quasinormal Modes, J. Cosmol. Astropart. Phys. 07, 066 (2020) [arXiv:2004.13061].
\bibitem{JL1} J. Liang, Quasinormal Modes of the Schwarzschild Black Hole Surrounded by the Quintessence Field in Rastall Gravity, Commun. Theor.
Phys. 70, 695 (2018).
15
\bibitem{YH} Y. Hu, C.-Y. Shao, Y.-J. Tan, C.-G. Shao, K. Lin, and W.-L. Qian, Scalar Quasinormal Modes of Nonlinear Charged Black Holes in
Rastall Gravity, EPL 128, 50006 (2020).
\bibitem{SG} S. Giri, H. Nandan, L. K. Joshi, and S. D. Maharaj, Geodesic Stability and Quasinormal Modes of Non-Commutative Schwarzschild
Black Hole Employing Lyapunov Exponent, Eur. Phys. J. Plus 137, 181 (2022).
\bibitem{DJ3} D. J. Gogoi, R. Karmakar, and U. D. Goswami, Quasinormal Modes of Non-Linearly Charged Black Holes Surrounded by a Cloud of
Strings in Rastall Gravity, arXiv:2111.00854 (2021).
\bibitem{AO} A. $\ddot{O}$vgu¨n, I˙. Sakallı and J. Saavedra, Quasinormal Modes of a Schwarzschild Black Hole Immersed in an Electromagnetic Universe,
Chin. Phys. C 42, no.10, 105102 (2018) arXiv:1708.08331[physics.gen-ph]].
\bibitem{AR} A. Rincon, P. A. Gonzalez, G. Panotopoulos, J. Saavedra and Y. Vasquez, Quasinormal modes for a non-minimally coupled scalar field
in a five-dimensional Einstein–Power– Maxwell background, Eur. Phys. J. Plus 137, no.11, 1278 (2022) [arXiv:2112.04793 [gr-qc]].
\bibitem{PA} P. A. Gonza´lez, A´ . Rinco´n, J. Saavedra and Y. Va´squez, Superradiant instability and charged scalar quasinormal modes for (2+1)-
dimensional Coulomb-like AdS black holes from nonlinear electrodynamics, Phys. Rev. D 104, no.8, 084047 (2021) [arXiv:2107.08611
[gr-qc]].
\bibitem{GP} G. Panotopoulos and A´ . Rinco´n, Quasinormal spectra of scale-dependent Schwarzschild–de Sitter black holes, Phys. Dark Univ. 31,
100743 (2021) [arXiv:2011.02860 [gr-qc]].
\bibitem{RG} R. G. Daghigh and M. D. Green, Validity of the WKB Approximation in Calculating the Asymptotic Quasinormal Modes of Black Holes,
Phys. Rev. D 85, 127501 (2012) [arXiv:1112.5397 [gr-qc]].
\bibitem{RG1} R. G. Daghigh and M. D. Green, Highly Real, Highly Damped, and Other Asymptotic Quasinormal Modes of Schwarzschild-Anti De
Sitter Black Holes, Class. Quant. Grav. 26, 125017 (2009) [arXiv:0808.1596 [gr-qc]].
\bibitem{AZ} A. Zhidenko, Quasinormal modes of Schwarzschild de Sitter black holes, Class. Quant. Grav. 21, 273-280 (2004) [arXiv:gr-qc/0307012
[gr-qc]].
\bibitem{AZ1} A. Zhidenko, Quasi-normal modes of the scalar hairy black hole, Class. Quant. Grav. 23, 3155-3164 (2006) [arXiv:gr-qc/0510039 [grqc]].
\bibitem{RA} R. A. Konoplya and A. Zhidenko, Quasinormal modes of black holes: From astrophysics to string theory, Rev. Mod. Phys. 83, 793-836
(2011) [arXiv:1102.4014 [gr-qc]].
\bibitem{YH}Y. Hatsuda, Quasinormal modes of black holes and Borel summation, Phys. Rev. D 101, no.2, 024008 (2020) [arXiv:1906.07232 [gr-qc]].
\bibitem{DS} D. S. Eniceicu and M. Reece, Quasinormal modes of charged fields in Reissner-Nordstr¨om backgrounds by Borel-Pad´e summation of
Bender-Wu series, Phys. Rev. D 102, no.4, 044015 (2020) [arXiv:1912.05553 [gr-qc]].
\bibitem{SL} S. Lepe and J. Saavedra, Quasinormal modes, superradiance and area spectrum for 2+1 acoustic black holes, Phys. Lett. B 617, 174-181
(2005) [arXiv:gr-qc/0410074 [gr-qc]].
\bibitem{MC} M. Chabab, H. El Moumni, S. Iraoui and K. Masmar, Phase Transition of Charged-AdS Black Holes and Quasinormal Modes : a Time
Domain Analysis, Astrophys. Space Sci. 362, no.10, 192 (2017) [arXiv:1701.00872 [hep-th]].
\bibitem{MC1} M. Chabab, H. El Moumni, S. Iraoui and K. Masmar, Behavior of quasinormal modes and high dimension RN–AdS black hole phase
transition, Eur. Phys. J. C 76, no.12, 676 (2016) [arXiv:1606.08524 [hep-th]].
\bibitem{MO} M. Okyay and A. O¨ vgu¨n, Nonlinear Electrodynamics Effects on the Black Hole Shadow, Deflection Angle, Quasinormal Modes and
Greybody Factors, J. Cosmol. Astropart. Phys. 2022, 009 (2022) [arXiv:2108.07766 [gr-qc]].
\bibitem{AO1} A. $\ddot{O}$ and K. Jusufi, Quasinormal Modes and Greybody Factors of f(R) Gravity Minimally Coupled to a Cloud of Strings in 2 + 1
Dimensions, Annals of Physics 395, 138 (2018) [arXiv:1801.02555 [gr-qc]].
\bibitem{RC2} R. C. Pantig, L. Mastrototaro, G. Lambiase and A. O¨ vgu¨n, Shadow, lensing, quasinormal modes, greybody bounds and neutrino propagation
by dyonic ModMax black holes, Eur. Phys. J. C 82, no.12, 1155 (2022) [arXiv:2208.06664[gr-qc]].
\bibitem{YY}Y. Yang, D. Liu, A. O¨ vgu¨n, Z. W. Long and Z. Xu, Quasinormal modes of Kerr-like black bounce spacetime, [arXiv:2205.07530[gr-qc]].
\bibitem{YY1} Y. Yang, D. Liu, A. O¨ vgu¨n, Z. W. Long and Z. Xu, Probing hairy black holes caused by gravitational decoupling using quasinormal
\bibitem{HAWKING}S. W. Hawking, Commun. Math. Phys. 43, 199 (1975b), [Erratum: Commun.Math.Phys. 46, 206 (1976)]
\bibitem{HH} H. Hassanabadi, W. S. Chung, B. C. L¨utf¨uo˘glu, and E. Maghsoodi, “Effects of a new extended uncertainty principle on Schwarzschild and Reissner–Nordstr¨om black holes thermodynamics,” Int. J. Mod. Phys. A 36, 2150036 (2021).
\bibitem{SH} S. Hassanabadi, J. Kˇr´ıˇz, W. S. Chung, B. C. L¨utf¨uo˘glu, E. Maghsoodi, and H. Hassanabadi, “Thermodynamics of the Schwarzschild and Reissner–Nordstr¨om black holes under higher-order generalized uncertainty principle,” Eur. Phys. J. Plus 136, 918 (2021),
arXiv:2110.01363 [gr-qc].
\bibitem{HC} Hao Chen, Bekir Can L¨utf¨uo˘glu, Hassan Hassanabadi, and Zheng-Wen Long, “Thermodynamics of the Reissner-Nordstr¨om black hole with quintessence matter on the EGUP framework,” Phys. Lett. B 827, 136994 (2022).
\bibitem{SW} Shao-Wen Wei Zhang, Yu-Peng and Yu-Xiao Liu, “Topological approach to derive the global Hawking temperature of (massive) BTZ black hole.” Physics Lett. B 810 (2020).
\bibitem{ALI} Ali $\ddot{O}$vg$\ddot{u}$n and Izzat Sakalli, “Hawking radiation via gaussbonnet theorem,” Ann. of Phys. 413, 168071 (2020).
\bibitem{SI} S. I Kruglov, “Magnetically charged black hole in framework of nonlinear electrodynamics model.” Int. J. of Mod. Phys. A 33 (2018).
\bibitem{SF} S. Fernando, “Greybody factors of charged dilaton black holes in 2 + 1 dimensions.” Gen. Relativ. Gravit. 37, 461481 (2005).
\bibitem{WK} Wontae Kim and John J. Oh, “Greybody factor and hawking radiation of charged dilatonic black holes,” J. of the Korean Phys. Society 52, 986–991 (2008).
\bibitem{JE} Jorge Escobedo, “Greybody factors,” Master’s Thesis, Uni. of Amsterdam 6 (2008).
\bibitem{MK} Maulik K. Parikh and Frank Wilczek, “Hawking radiation as tunneling,” Phys. Rev. Lett. 85, 5042–5045 (2000).
\bibitem{CH} Chris H. Fleming, “Hawking radiation as tunneling,” Uni. of Maryland. Dept. of Phys., Tech. Rep (2005).
\bibitem{GB}Matt Visser, “Some general bounds for one-dimensional scattering,” Phys. Rev. A 59, 427–438 (1999).
\bibitem{GB1}Petarpa Boonserm and Matt Visser, “Bounding the bogoliubov coefficients,” Annals of Physics 323, 2779–2798 (2008).
\bibitem{WJ} W. Javed, I. Hussain, and A. O¨ vgu¨n, “”Weak deflection angle of KazakovSolodukhin black hole in plasma medium using GaussBonnet theorem and its greybody bonding.” Eur. Phys. J. Plus 137 (2022).
\bibitem{SIMPSON}A. Simpson and M. Visser, JCAP 02, 042 (2019).
\bibitem{kirill} "Field sources for Simpson-Visser spacetimes", Kirill A. Bronnikov and Rahul Kumar Walia, Phys. Rev. D 105, 044039.
\bibitem{SIMPSON1}A. Simpson, P. Martin-Moruno and M. Visser, Class. Quant. Grav. 36, 145007 (2019).
\bibitem{LOBO}F. S. N. Lobo, M. E. Rodrigues, M. V. d. S. Silva, A. Simpson and M. Visser, [arXiv:2009.12057 [gr-qc]].
\bibitem{schutz} Schutz B F and Will C M 1985 BLACK HOLE NORMAL MODES: A SEMIANALYTIC APPROACH \emph{Astrophys. J. Lett.} \textbf{291} L33-L36
\bibitem{iyer} Iyer S and Will C M 1987 Black Hole Normal Modes: A {WKB} Approach. 1. Foundations and Application of a Higher Order {WKB} Analysis of Potential Barrier Scattering \emph{Phys. Rev. D} \textbf{35} 3621
\bibitem{iyer1}
Iyer S 1987 BLACK HOLE NORMAL MODES: A WKB APPROACH. 2. SCHWARZSCHILD BLACK HOLES
\emph{Phys. Rev. D} \textbf{35} 3632
\bibitem{konoplya1}
Konoplya R A 2003
Quasinormal behavior of the d-dimensional Schwarzschild black hole and higher order WKB approach
\emph{Phys. Rev. D} \textbf{68} 024018
\bibitem{jerzy}Jerzy Matyjasek and Micha l Opala, “Quasinormal modes of black holes: The improved semianalytic approach,”
Phys. Rev. D 96, 024011 (2017).
\bibitem{konoplya2} RA Konoplya, A Zhidenko, and AF Zinhailo, “Higher order wkb formula for quasinormal modes and grey-body factors:
recipes for quick and accurate calculations,” Classical and Quantum Gravity 36, 155002 (2019).
\bibitem{KD}K. Destounis, R. P. Macedo, E. Berti, V. Cardoso and J. L. Jaramillo, “Pseudospectrum of
Reissner-Nordström black holes: quasinormal mode instability and universality,” [arXiv:2107.09673
[gr-qc]].
\bibitem{OVG}M. Okyay and A. $\ddot{O}$vg$\ddot{u}$n, Nonlinear electrodynamics effects on the black hole shadow, deflection angle, quasinormal modes and greybody factors, JCAP 01 (01), 009, arXiv:2108.07766 [gr-qc].
\bibitem{GB2}Boonserm. P, “Rigorous bounds on transmission, reflection and bogoliubov coefficients,” Ph.D. thesis, Victoria Uni. Wellington (2009).
\bibitem{BEK} Black hole thermodynamics, Physics Today 24, Bekenstein 1980.
\bibitem{KEIF} Classical and Quantum black holes by Keifer 1999.
\end{thebibliography}
\end{document}